\documentclass[pre,preprint]{revtex4}
\usepackage{graphicx,graphics}
\usepackage{mathrsfs}
\usepackage{color}
 
\begin{document}  
 
\title{Synchronization Stability of Coupled Near-Identical Oscillator Network}

\author{Jie Sun}
\email{sunj@clarkson.edu} 
\affiliation{Department of Mathematics \& Computer Science, Clarkson University, Potsdam, NY 13699-5815, USA}
\author{Erik M.~Bollt}
\email{bolltem@clarkson.edu}
\affiliation{Department of Mathematics \& Computer Science, Clarkson University, Potsdam, NY 13699-5815, USA}
\author{Takashi Nishikawa}
\email{tnishika@clarkson.edu}
\affiliation{Department of Mathematics \& Computer Science, Clarkson University, Potsdam, NY 13699-5815, USA}

\begin{abstract}
We derive variational equations to analyze the stability of synchronization for coupled near-identical oscillators. To study the effect of parameter mismatch on the stability in a general fashion, we define master stability equations and associated master stability functions, which are independent of the network structure. In particular, we present several examples of coupled near-identical Lorenz systems configured in small networks (a ring graph and sequence networks) with a fixed parameter mismatch and a large Barabasi-Albert scale-free network with random parameter mismatch. We find that several different network architectures permit similar results despite various mismatch patterns.
\end{abstract}

\maketitle

\section{Introduction}
The phenomena of synchronization has been found in various aspects of nature and science\cite{STROGATZ_BOOK03}. Its applications have ranged widely from biology\cite{STROGATZ_SCIAM93, BUONO_JMATHBIOL01} to mathematical epidemiology\cite{He_BIOLSCI03}, and chaotic oscillators\cite{PECORA_PRL90}, to communicational devices in engineering\cite{CUOMO_PRL93}, etc. With the development of theory and application in complex networks\cite{ALBERT_RMP02}, the study of synchronization between a large number of coupled dynamically driven oscillators has become a popular and exciting developing topic, see for example \cite{BOCCALETTI_PHYSREP02, NISHIKAWA_PRL03, Li_IEEE03, SKUFCA_MBE04, STILWELL_SIADS06, ARENAS_PRL06}.

To model the coupled dynamics on a network (assumed to be unweighted and undirected and connected throughout this paper), we consider, for $i = 1,2,..., N$:
\begin{equation}\label{origdyn}
	\dot{w_i} = f(w_i,\mu_i) - g\sum_{j=1}^{N}{L_{ij}H(w_j)}
\end{equation}
where
$w_i\in \Re^{m}$ is used to represent the dynamical variable on the $i$th unit; 
$f:\Re^{m}\times\Re^{p}\rightarrow \Re^{m}$ is the individual dynamics (usually chaotic dynamics for most interesting problems) on $i$ and
$\mu_i\in \Re^{p}$ is the corresponding parameter;
$L\in \Re^{N\times N}$ is the graph laplacian defined by $L_{ij}\equiv-1$ if there is an edge connecting node $i$ and $j$ and the diagonal element $L_{ii}$ is defined to be the total number of edges incident to node $i$ in the network;
$H:\Re^{m}\rightarrow\Re^{m}$ is a uniform coupling function on the net;
and $g\in \Re$ is the uniform coupling strength (usually $>0$ for diffusive coupling). The whole system can be represented compactly with the use of Kronecker product:
\begin{eqnarray}
	\mbox{{\boldmath $\dot w$}} = \mbox{{\boldmath $f(w,\mu)$}} - g\cdot L\otimes H(\mbox{{\boldmath $w$}})
\end{eqnarray}
where {\boldmath $w$}$=(w_1^T,w_2^T,...w_N^T)^T$ is a column vector of all the dynamic variables, and likewise for {\boldmath $\mu$} and {\boldmath $f$}; and $\otimes$ is the usual Kronecker product\cite{LANCASTER_BOOK}.

The majority of the theoretical work has been focused on {\it identical synchronization} where $\max_{i,j}{||w_i(t)-w_j(t)||}\rightarrow 0$ as $t\rightarrow\infty$, since it is in this situation the stability analysis can be carried forward by using the master stability functions proposed in the seminal work \cite{PECORA_PRL98}. However, realistically it is impossible to find or construct a coupled dynamical system made up of exactly identical units, in which case identical synchronization rarely happens, but instead, a {\it nearly synchronous state} often takes place instead, where $\max_{i,j}{||w_i(t)-w_j(t)||}\leq C$ for some small constant $C>0$ as $t\rightarrow\infty$.

It is thus important to analyze how systems such as Eq.~(\ref{origdyn}) evolve, when parameter mismatch appears. In \cite{RESTREPO_PRE04}, similar variational equations were used to study the impact of parameter mismatch on the possible de-synchronization. To study the effect of parameter mismatch on the stability of synchronization, and more specifically, to find the distance bound $C$ in terms of the given parameters in Eq.~(\ref{origdyn}), we derive variational equations of system such as Eq.~(\ref{origdyn}) and extend the master stability function approach to this case, to decompose the problem into two parts that depend on the individual dynamics and network structure respectively.

\section{Theory: Master Stability Equations and Functions}
\subsection{Derivation of Variational Equations}
When the parameters $\mu_i$ of individual units in Eq.~(\ref{origdyn}) are close to each other, centered around their mean $\bar{\mu}$, the coupled units $w_i$ are found empirically to satisfy $\max_{i,j}{||w_i(t)-w_j(t)||}\leq C$ for some $C>0$ as $t\rightarrow\infty$, referred to as {\it near synchronization}\cite{RESTREPO_PRE04}. When such near synchronization state exists, the {\it average trajectory} well represents the collective behavior of all the units. The average trajectory $\bar{w}\equiv \frac{1}{N}\sum_{i=1}^{N}{w_i}$ of Eq.~(\ref{origdyn}) satisfies:
\begin{eqnarray}
	\dot{\bar{w}} &=& \frac{1}{N}\sum_{i=1}^{N}{f(w_i,\mu_i)} - g\sum_{i=1}^{N}\sum_{j=1}^{N}{L_{ij}H(w_j)} \nonumber\\
					&=& \frac{1}{N}\sum_{i=1}^{N}{f(w_i,\mu_i)},
\end{eqnarray}
since $\sum_{i=1}^{N}{L_{ij}} = 0$ by the definition of $L$. The variation $\eta_i \equiv w_i - \bar{w}$ of each individual unit is found to satisfy the following {\it variational equation}:
\begin{eqnarray}
	\dot{\eta_i} = D_{w}f(\bar{w},\bar{\mu}) \eta_i  - g\sum_{j=1}^{N}{L_{ij}DH(\bar{w}) \eta_j} 
	 + D_{\mu}f(\bar{w},\bar{\mu}) \delta\mu_i,
\end{eqnarray}
where $\bar{\mu}\equiv\sum_{i=1}^{N}{\mu_i}$ and $\delta\mu_i\equiv\mu_i-\bar{\mu}$; and $D_{w}$ represents the derivative matrix with respect to $w$ and likewise for $D_{\mu}$ and $DH$. The above variational equations can be represented in Kronecker product form as:
\begin{eqnarray}\label{vareq}
	\mbox{{\boldmath $\dot\eta$}} = \Big[ I_N\otimes D_{w}f - g\cdot L \otimes DH \Big]\mbox{{\boldmath $\eta$}} + 
	\Big[ I_N \otimes D_{\mu}f \Big]\mbox{{\boldmath $\delta\mu$}},
\end{eqnarray}
where $\eta_i$ are stacked into a column vector $\mbox{\boldmath $\eta$}$ and likewise for $\delta\mu_i$.

\subsection{Decomposition of the Variational Equations}
Since we are dealing with undirected graph, the associated $L$ is symmetric and positive semi-definite, and thus $L$ is diagonalizable: $L = P\Lambda P^{T}$\cite{LANCASTER_BOOK}, where $\Lambda$ is the diagonal matrix whose $i$th diagonal entry $\lambda_i$ is the $i$th eigenvalue of $L$ (arranged in the order $\lambda_1\leq\lambda_2\leq...\leq \lambda_N$); and $P$ is the orthogonal matrix whose $i$th column $v_i=(v_{1,i},...,v_{N,i})^T$ is the normalized eigenvector associated with $\lambda_i$, and all these $v_i$ form an orthonormal basis of $\Re^{N}$. Note that because of $\sum_{i=1}^{N}{L_{ij}} = 0$, we always have $\lambda_1=0$ with $v_1=\frac{1}{\sqrt{N}}(1,...,1)^T$; and since we have assumed that the graph is connected, the following holds: 
$\lambda_1\equiv 0 < \lambda_2 \leq ... \leq \lambda_N$.

We may uncouple the variational equation Eq.~(\ref{vareq}) by making the change of variables 
\begin{equation}
	\mbox{{\boldmath $\zeta$}} \equiv (P^{T}\otimes I_m) \mbox{{\boldmath $\eta$}},
\end{equation}
or more explicitly, for each $i$, 
\begin{equation}\label{etazeta}
	\zeta_i \equiv v_{1,i}\eta_1 + v_{2,i}\eta_2 + ... + v_{N,i}\eta_N,
\end{equation}
to yield:
\begin{eqnarray}\label{vareq2}
	\mbox{{\boldmath $\dot\zeta$}} = \Big[ I_N \otimes D_{w}f - g\cdot \Lambda \otimes DH \Big] \mbox{{\boldmath $\zeta$}}
	+ \Big[ P^{T} \otimes D_{\mu}f \Big] \mbox{{\boldmath $\delta\mu$}}.
\end{eqnarray}
where $\mbox{\boldmath $\zeta$}\equiv(\zeta_1^T,...,\zeta_N^T)^T.$ Note that since 
$\sum_{i=1}^{N}{\eta_i} \equiv \sum_{i=1}^{N}{(w_i-\bar{w})} = 0$, 
and $v_1=\frac{1}{\sqrt{N}}(1,...,1)^T$, the following holds: $\zeta_1\equiv 0$, by Eq.~(\ref{etazeta}).

Note that since the transformation {\boldmath$\zeta \equiv $} $(P^{T}\otimes I_m)${\boldmath$\eta$} is an orthogonal transformation, $||\mbox{\boldmath$\zeta$}|| \equiv ||\mbox{\boldmath$\eta$}||$ with the choice of Euclidean norm. In other words, for $||.||$ being the usual Euclidean distance, we have:
\begin{equation}
	\sum_{i=1}^{N}{||\zeta_{i}||^2} \equiv \sum_{i=1}^{N}{||\eta_{i}||^2}.
\end{equation}

The homogeneous part in Eq.~(\ref{vareq2}) has block diagonal structure and we may write for each eigenmode ($i=2,3,...,N$):
\begin{eqnarray}\label{vareq3}
	\dot{\zeta_i} = \Big[D_{w}f - g\lambda_i DH \Big]\zeta_i 
		+ D_{\mu}f \cdot \Big(\sum_{j=1}^{N}{v_{j,i}\delta\mu_j} \Big).
\end{eqnarray}
The vector $\sum_{j=1}^{N}{v_{j,i}\delta\mu_j}$ is the weighted average of parameter mismatch vectors, weighted by the eigenvector components associated with $\lambda_i$, and may be thought of as the length of projection of the parameter mismatch vector onto the eigenvector $v_i$.

\subsection{Extended Master Stability Equations and Functions}
The variational equation in the new coordinate system Eq.~(\ref{vareq3}) suggests a generic approach\cite{PECORA_PRL98} to study the stability of synchronization for a given network coupled dynamical system investigating on the effect of $\lambda_i$ and $\Big(\sum_{j=1}^{N}{v_{j,i}\delta\mu_j} \Big)$ on the solution of Eq.~(\ref{vareq3}). We define an {\it extended master stability equation} 
\footnote{Note here that to obtain the MSF based on Eq.~(\ref{homoxi}), we need the actual average trajectory $\bar{w}(t)$, which can only be obtained by solving the whole system Eq.~(\ref{origdyn}). However, we found that the trajectory solved from a single system $\dot{s} = f(s,\bar{\mu})$ could be used instead, resulting in good approximation of $\Omega$. The supporting work for proving the shadowability of $\bar{w}$ by $s$ will be reported elsewhere. }
for near identical coupled dynamical systems:
\begin{equation}\label{EMSE}
	\dot{\xi} = \Big[ D_{w}f - \alpha\cdot DH \Big] + D_{\mu}f\cdot\psi
\end{equation}
where we have introduced two auxiliary parameters, $\alpha\in\Re$ and $\psi\in\Re^{p}$. This generic equation decomposes the stability problem into two separate parts: one that depends only on the individual dynamics and the coupling function, and one that depends only on the graph Laplacian and parameter mismatch. Note that the latter not only depends on the spectrum of $L$ as in \cite{PECORA_PRL98}, but also on the combination of the eigenvectors and parameter mismatch vector. 

Once the stability of Eq.~(\ref{EMSE}) is determined as a function of $\alpha$ and $\psi$, the stability of any coupled network oscillators as described by Eq.~(\ref{origdyn}), for the given $f$ and $H$ used in Eq.~(\ref{EMSE}), can be found by simply setting 
\begin{equation}\label{setalpha}
	\alpha = g\lambda_i
\end{equation}
and 
\begin{equation}\label{setpsi}
	\psi = \sum_{j=1}^{N}v_{j,i}\delta\mu_j
\end{equation}
where $\lambda_i, v_{j,i}, \delta\mu_j$ can be obtained by the knowledge of the underlying network structure $L$ and parameter mismatch pattern. Thus, we have reduced the stability analysis of the original $mN$-dimensional problem to that of an $m$-dimensional problem with one additional parameter, combined with an eigen-problem. 

The associated {\it master stability function} (MSF) $\Omega(\alpha,\psi)$ of Eq.~(\ref{EMSE}) is defined as:
\begin{equation}\label{EMSF}
	\Omega(\alpha,\psi) \equiv \lim_{T\rightarrow\infty}{ \sqrt{ \frac{1}{T} \int_{0}^{T}{||\xi(t)||^2dt} }     }
\end{equation}
when the limit exists, where $\xi$ is a solution of Eq.~(\ref{EMSE}) for the given $(\alpha,\psi)$ pair.

For a given coupled oscillator network by Eq.~(\ref{origdyn}), we have the following equation, based on the generic MSF $\Omega$:
\begin{eqnarray}
	&&\lim_{T\rightarrow\infty}{ \sqrt{ \frac{1}{T} \int_{0}^{T}{\sum_{i=1}^{N}{||w_i(t)-\bar{w_i}(t)||^2}dt} }     } \nonumber\\
	&\equiv&   \lim_{T\rightarrow\infty}{ \sqrt{ \frac{1}{T} \int_{0}^{T}{\sum_{i=1}^{N}{||\eta_i(t)||^2}dt} }     } 
	\equiv    \lim_{T\rightarrow\infty}{ \sqrt{ \frac{1}{T} \int_{0}^{T}{\sum_{i=2}^{N}{||\zeta_i(t)||^2}dt} }     } \nonumber\\
	& = & \sum_{i=2}^{N}{ \Omega^2(g\lambda_i,\psi_i)}
\end{eqnarray}
where $\lambda_i$ are the eigenvalues of the graph Laplacian and $\psi_i$ is obtained through Eq.~(\ref{setpsi}).
Thus, once the MSF for the dynamics $f$ and coupling function $H$ has been computed, it can be used to compute the asymptotic total distance from single units to the average trajectory: 
$<\sum_{i=1}^{N}{||w_i(t)-\bar{w_i}(t)||^2}>$
\footnote{Notation $<a(t)>$ is introduced and used throughout, to represent the {\it asymptotic root mean square}: 
$\sqrt{ \lim_{T\rightarrow\infty}{ \frac{1}{T} \int_{0}^{T}{a(t)dt}  } }$ 
for the trajectory $a(t)$.}
for any coupled oscillator network by summing up the corresponding $\Omega^2(g\lambda_i,\psi_i)$ and take the square root.

\begin{figure}[ht]
\centering
\includegraphics*[width=1\textwidth]{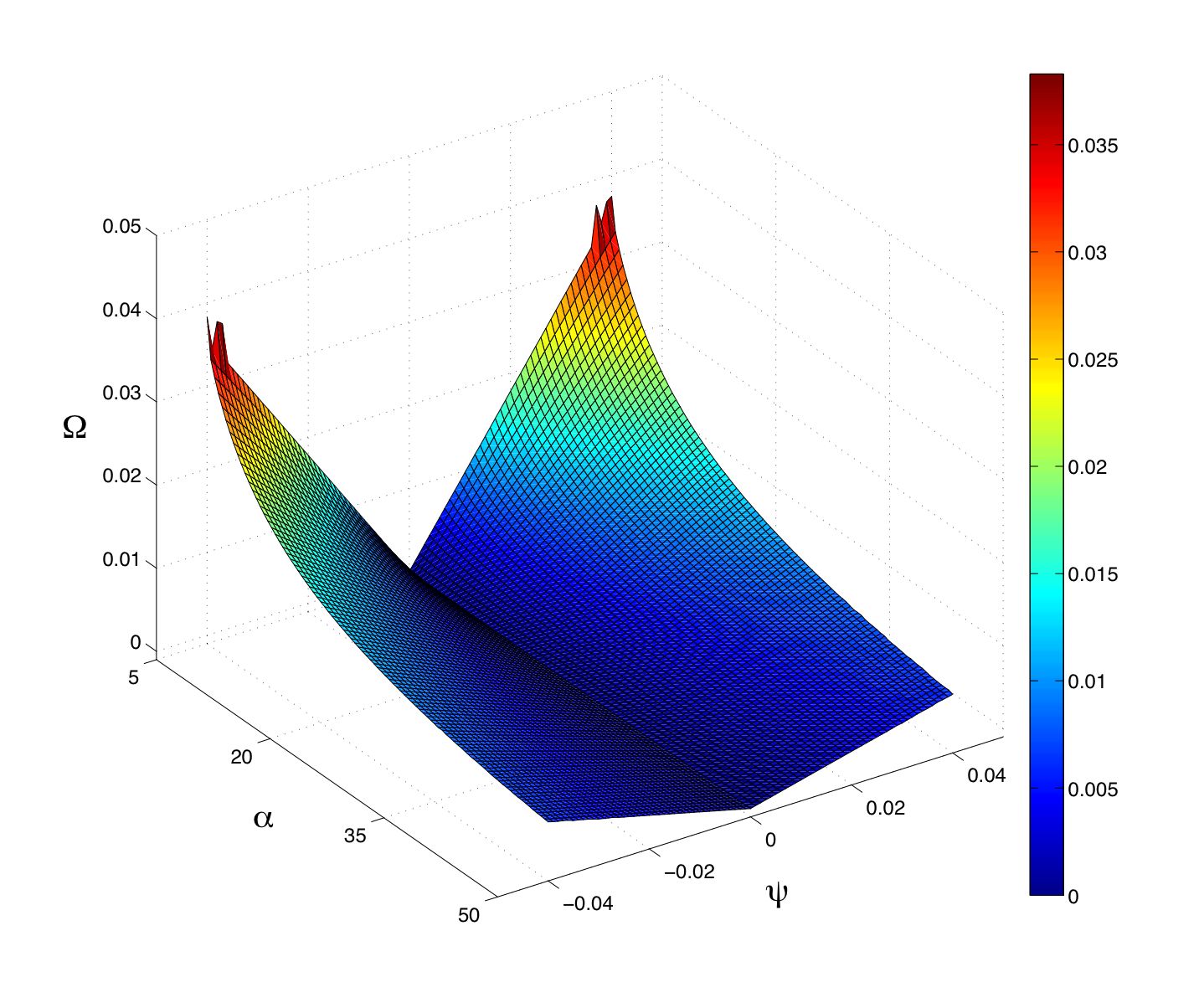}
\caption{(Color online) MSF for Lorenz system Eq.~(\ref{lorenz}), with $\sigma=10,\beta=\frac{8}{3}$, adjustable parameter being $r$, and coupling function $H$ being an identity matrix operator. The domain shown here is for $\alpha$ from $5$ to $50$ and $\psi$ from $-0.04$ to $0.04$, while the actual valid domain of MSF could be as large as the stability region of the identical Lorenz system (in this case is the region $\alpha > \lambda_0$ where $\lambda_0$ is the largest lyapunov exponent of the original Lorenz system ($\approx 1$)). }
\label{MSFpic}
\end{figure}

In Fig.~\ref{MSFpic} we plot the MSF for $f$ being Lorenz equations:
\begin{eqnarray}\label{lorenz}
	\dot{x} &=& \sigma(y-x) \nonumber\\
	\dot{y} &=& x(r-z) - y        \nonumber\\
	\dot{z} &=& xy - \beta z
\end{eqnarray}
as individual dynamics in Eq.~(\ref{origdyn}) ($w=[x,y,z]^T$). The parameters are chosen as: $\sigma=10, \beta=\frac{8}{3}$, and $r$ is allowed to be adjustable, i.e., $r$ is the $\mu$ in Eq.~(\ref{origdyn}). The coupling function $H$ is taken as: $H(w)=w$, i.e., an identity matrix operator.

\subsection{Conditions for Stable Near Synchronization}
For near synchronization to appear in the presence of parameter mismatch, it is required that the system described by Eq.~(\ref{origdyn}) in the absence of parameter mismatch undergoes stable identical synchronization, which can be checked by using MSF\cite{PECORA_PRL98}. In this case, the largest Lyapunov exponent of the synchronous trajectory associated with the homogeneous variational equation:
\begin{equation}\label{homoxi}
	\dot{\xi} = \Big[ D_{w}f - \alpha\cdot DH \Big] \xi
\end{equation}
is negative, and its solution can be written as $\xi^{*}(t) = \Phi(t,0)\xi(0)$ where $\Phi(t,\tau)$ is the fundamental transition matrix
\footnote{This transition matrix, as a function of two time variables $t$ and $\tau$, can be obtained by the {\it Peano-Baker series} as long as $D_{w}f - \alpha\cdot DH$ is continuous. See \cite{RUGH_BOOK}, Ch.3.}, satisfying
\begin{equation}\label{bdPhi}
	||\Phi(t,\tau)|| \leq \gamma e^{-\lambda(t-\tau)}
\end{equation}
for $t\geq\tau$ and some finite positive constants $\gamma$ and $\lambda$. 

Note that the transition to loss of stability at certain time instances can occur due to the embedded periodic orbits\cite{VENK_PRL96, RESTREPO_PRE04}, in which case the above inequality will not hold. In this paper we consider the situation where Eq.~(\ref{bdPhi}) holds for most of the time, with $\lambda$ being the Lyapunov exponent of the trajectory associated with Eq.~(\ref{homoxi}), although at certain time instances Eq.~(\ref{bdPhi}) need not hold, as discussed in \cite{VENK_PRL96, RESTREPO_PRE04}, referred to as bubbling transition\cite{VENK_PRE96}.

The solution to Eq.~(\ref{EMSE}) can be expressed as:
\begin{equation}\label{solxi}
	\xi(t) = \Phi(t,0)\xi(0) + \int_{0}^{t}{\Phi(t,\tau)b(\tau)d\tau},
\end{equation}
where we have defined $b(\tau) \equiv D_{\mu}f(s(\tau),\bar{\mu})\cdot\psi$. Under the condition of Eq.~(\ref{bdPhi}), we have the following bound for $\xi(t)$:
\begin{eqnarray}
	||\xi(t)|| &\leq& ||\Phi(t,0)||\cdot||\xi(0)|| + \int_{0}^{t}{||\Phi(t,\tau)||d\tau}\cdot\sup_{t}||{b(t)}|| \nonumber\\
	                    &\leq& \gamma e^{-\lambda t}||\xi(0)|| + \frac{\gamma}{\lambda}(1-e^{-\lambda t})\sup_{t}||{b(t)}|| \nonumber\\
	                    &\rightarrow& \frac{\gamma}{\lambda}\sup_{t}||{b(t)}||  \mbox{ }\mbox{ as } t\rightarrow\infty.
\end{eqnarray}
Thus, the conditions for stable near synchronization of Eq.~(\ref{origdyn}) are: 
\begin{enumerate}
	\item  The corresponding identical system (without parameter mismatch) is stably synchronized, or equivalently, the associated variational equation Eq.~(\ref{homoxi}) is exponentially stable;
	\item The inhomogeneous part $b(\tau) \equiv D_{\mu}f(s(\tau),\bar{\mu})\cdot\psi$ in Eq.~(\ref{EMSE}) is bounded.
\end{enumerate}
These conditions are sufficient to guarantee the boundness of pairwise distance between any two units, so that near synchronous state is stable.

Eq.~(\ref{bdPhi}) and Eq.~(\ref{solxi}) also allow us to analyze quantitatively the magnitude of asymptotic error of a near-identical system such as Eq.~(\ref{origdyn}). For all other variables being the same, if the magnitude of parameter mismatch is scaled by a factor $k$, then the corresponding variation will become:
\begin{equation}\label{solxi2}
	\widetilde\xi(t) = \Phi(t,0)\xi(0) + k \cdot \int_{0}^{t}{\Phi(t,\tau)b(\tau)d\tau}
\end{equation}
where $\xi(t)$ denotes the variation of the original unscaled near-identical system, which follows Eq.~(\ref{solxi}). The first term of both Eq.~(\ref{solxi}) and Eq.~(\ref{solxi2}) goes to zero according to Eq.~(\ref{bdPhi}), so that asymptotically the following holds: $\widetilde{\xi}(t) = k\cdot\xi(t)$, i.e., the variation is scaled by the same factor correspondingly.

\section{Examples of Application}

\subsection{Methodology}
When the units coupling through the network are known exactly, meaning that the parameter of each unit is known, then from Eq.~(\ref{setalpha}) and Eq.~(\ref{setpsi}) we may use the $\Omega$ obtained from MSF at  the corresponding $(\alpha,\psi)$ pairs. In Sec. 3.2 and Sec. 3.3 we illustrate this with examples of small networks.

On the other hand, for large networks, in the case that parameters of individual units are not known exactly, but follow a Gaussian distribution:   $\delta\mu_i\sim N(\bar{\mu},\epsilon^2)$, then in Eq.~(\ref{vareq3}) we have: 
\begin{eqnarray}
	\Big( \sum_{j=1}^{N}{v_{j,i}\delta\mu_j} \Big) &\sim& N(\bar{\mu},\sum_{j=1}^{N}{v_{j,i}^2}\epsilon^2) \nonumber\\
	&\sim& N(\bar{\mu},\epsilon^2)
\end{eqnarray}
assuming the $\delta\mu_i$ are identical and independent. The standard deviation $\epsilon$ may be used, as an expected bound for $\psi$ in Eq.~(\ref{setpsi}), to compute an {\it expected MSF} to predict the possible variation of individual units to the average trajectory. In Sec. 3.3 a scale-free network with $N=500$ vertices is used to illustrate.

In all the examples, the individual dynamics is the Lorenz equation Eq.~(\ref{lorenz}), with parameters $\sigma=10, \beta=\frac{8}{3}$, and $r_i=28+\delta r_i$ where $\delta r_i$ is the parameter mismatch on unit $i$. The coupling function is chosen as $H(w)=w$ with coupling strength $g$ specified differently in each example. The variation of individual units to the average trajectory $<\sum_{i=1}^{N}{||\eta_{i}(t)||^2}>$ is approximated by $T=200$ with equally time spacing $\tau=0.01$.

\subsection{Example: Ring Graph}
\begin{figure}
\centering
\includegraphics*[width=0.3\textwidth]{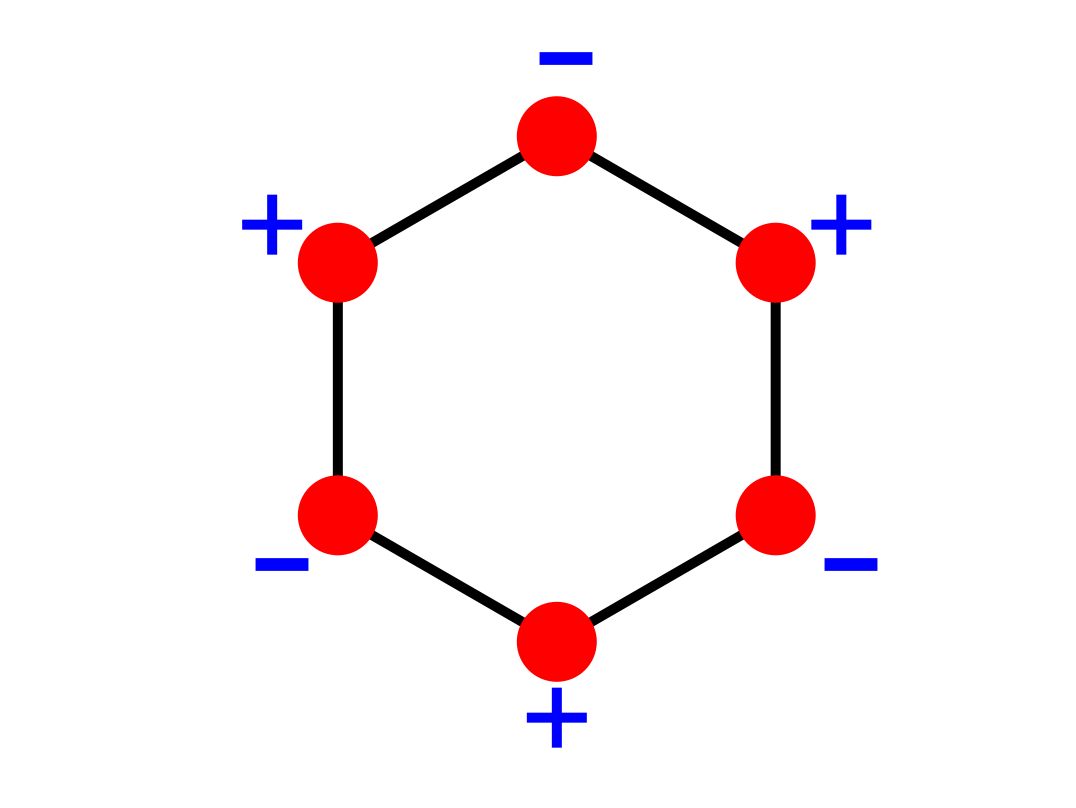}
\includegraphics*[width=0.3\textwidth]{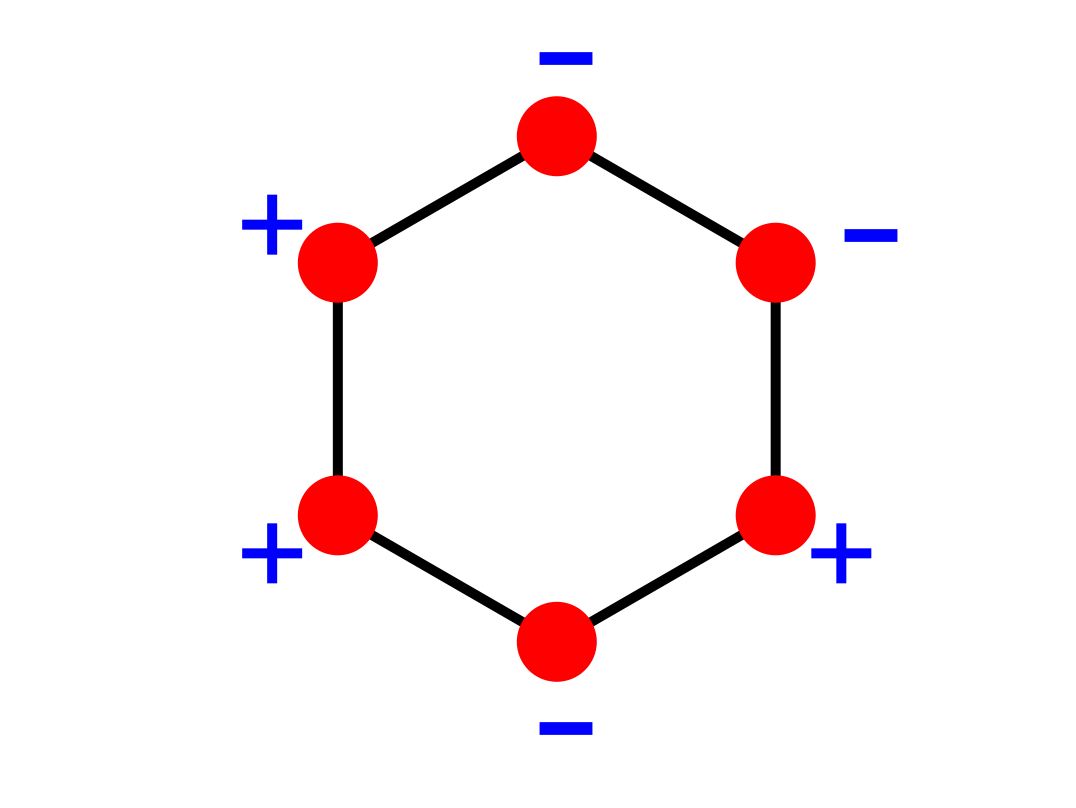}
\includegraphics*[width=0.3\textwidth]{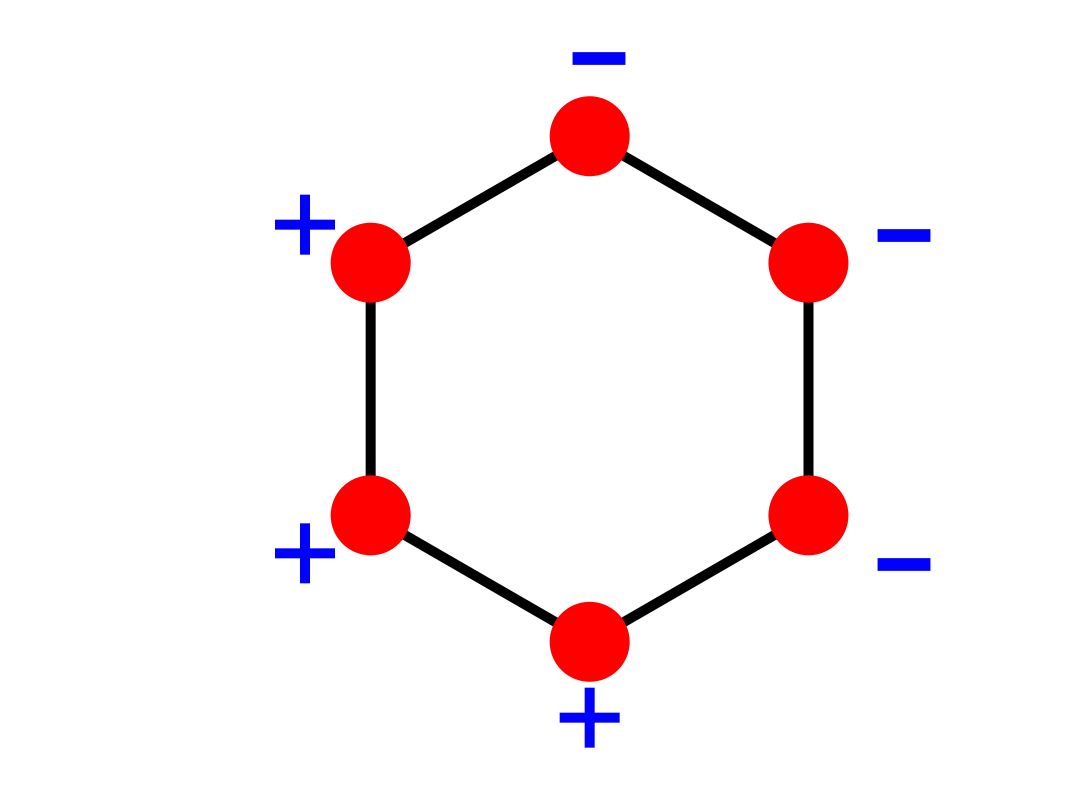}
\caption{(Color online) Ring graph (red circles represent vertices and black lines represent edges) with specific parameter mismatch on each unit. The magnitude of parameter mismatch on each unit is assumed to be the same, $\epsilon$. The plus/minus sign on a vertex represents the corresponding sign of mismatch on that unit, $"+"$ for $+\epsilon$ and $"-"$ for $-\epsilon$. So the left graph has the parameter mismatch pattern (starting from the top unit): 
$[-\epsilon,+\epsilon,-\epsilon,+\epsilon,-\epsilon,+\epsilon]$, 
the middle graph has the pattern 
$[-\epsilon,-\epsilon,+\epsilon,-\epsilon,+\epsilon,+\epsilon]$, 
and the right graph has the pattern 
$[-\epsilon,-\epsilon,-\epsilon,+\epsilon,+\epsilon,+\epsilon]$.
}
\label{ringgraph}
\end{figure}

\begin{figure}
\centering
\includegraphics*[width=0.7\textwidth]{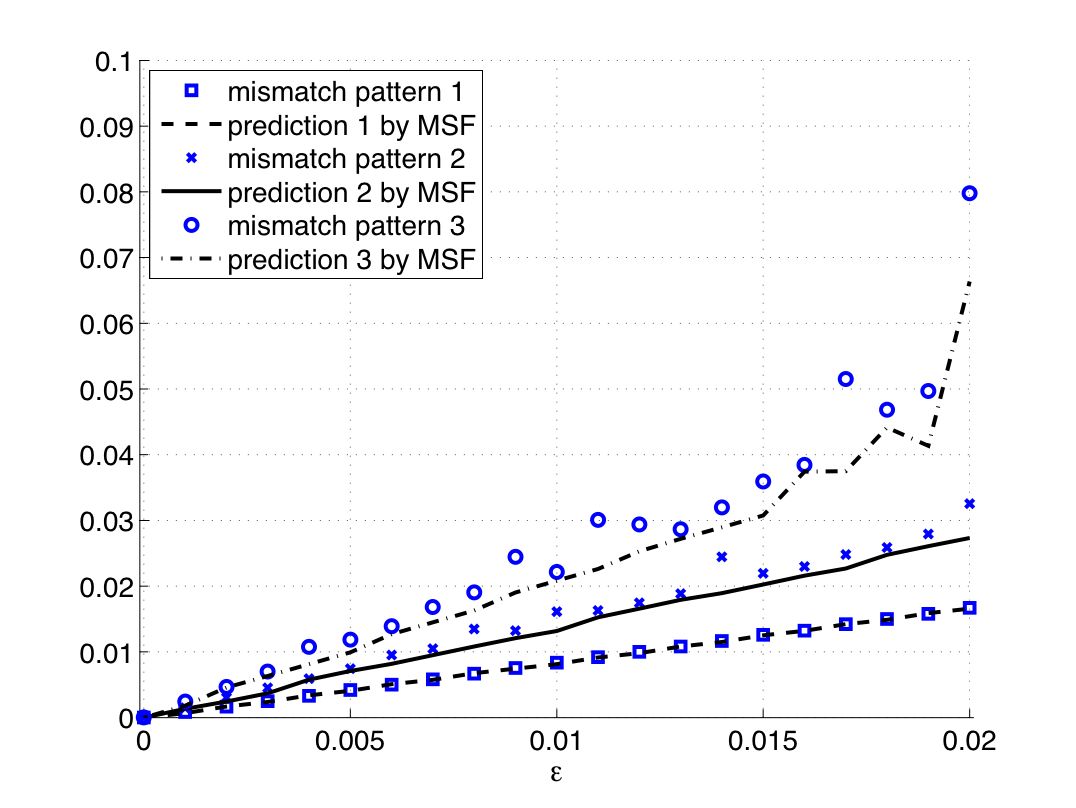}
\caption{(Color online) Validating MSF on a ring graph. Here the coupling strength is $g=5$. The units are coupled through a ring graph, with specific parameter mismatch patterns as shown in Fig.~\ref{ringgraph}. The vertical axis represents the average variation at each given $\epsilon$. Blue squares, crosses, and circles are obtained from actual time series, computed through
$<\sum_{i=1}^{N}{||\eta_{i}(t)||^2}>$
where $\eta_i(t)$ is the distance from unit $i$ to the average trajectory at time $t$.
Black lines (dashed, solid, and dotted) are theoretical prediction 
$\sqrt{ \sum_{i=1}^{N}{\Omega^2(\alpha_i,\psi_i)} }$
from MSF at $(\alpha_i,\psi_i)$ paris, where $(\alpha_i,\psi)$ are computed according to Eq.~(\ref{setalpha}) and Eq.~(\ref{setpsi}).
}
\label{ringvar}
\end{figure}

We consider a small and simple graph to illustrate. The graph as well as three different patterns of parameter mismatch are shown in Fig.~\ref{ringgraph}. In Fig.~\ref{ringvar} we show the actual variation on individual units and that by MSF. 

The MSF predicts well the actual variations found in this near-identical oscillator network, in all three cases. Furthermore, the way parameter mismatch are distributed in the graph is relevant, as a consequence of Eq.~(\ref{vareq3}). From left to right in Fig.~(\ref{ringgraph}), the parameter mismatch is distributed more heterogeneously, resulting in larger variation along the near synchronous trajectory.

\subsection{Example: Sequence Networks}
\begin{figure}
\centering
\includegraphics*[width=0.8\textwidth]{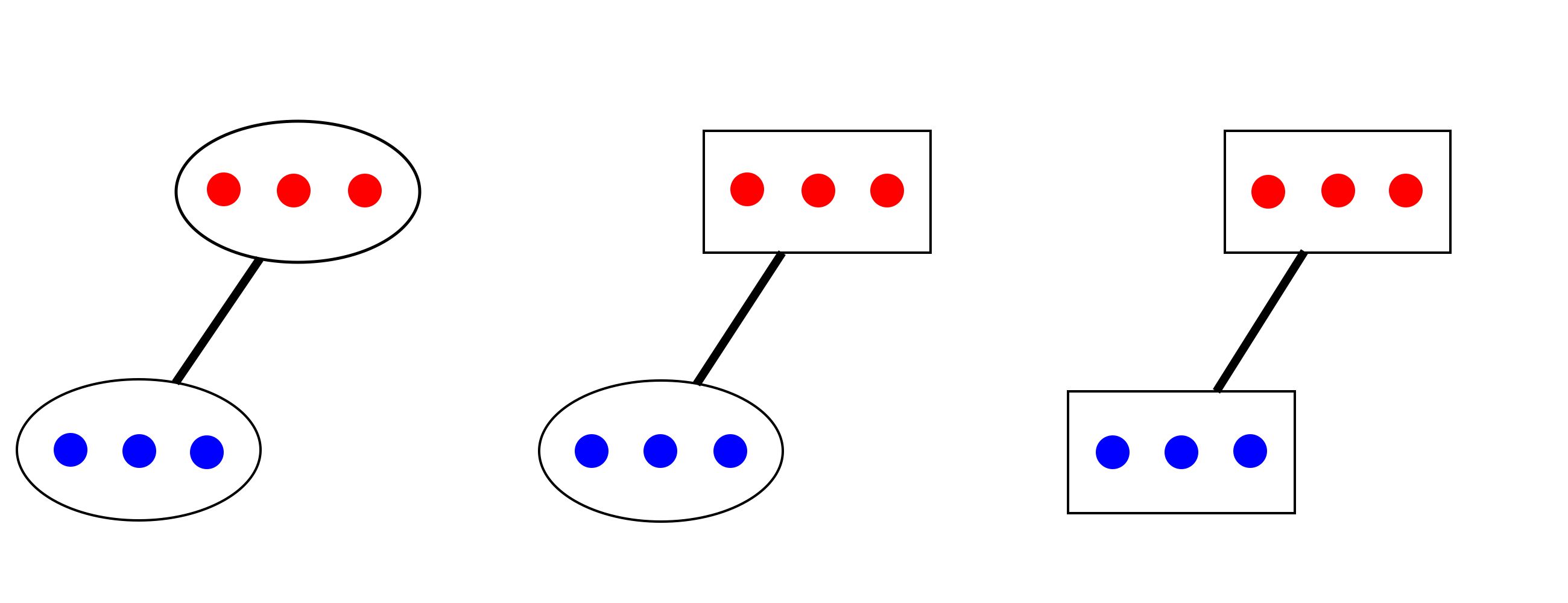}
\caption{(Color online) 2-letter sequence nets\cite{SUN_PRE08} consisting of $6$ vertices and $2$ layers  (red and blue circles represent vertices and black lines represent edges) obtained from the same creation sequence $(A,A,A,B,B,B)$ under three different rules, on the left the connection rule is $B\rightarrow A$, meaning that whenever a vertex of type $B$ is added into the current net, it connects to all previous vertices of type $A$, thus a bipartite complete graph is created based on this sequence; on the middle the connection rule is $B\rightarrow A,B$, resulting in a threshold graph; while on the right the rule $A\rightarrow A,B; B\rightarrow A,B$ is applied to yield a complete graph. Ovals and boxes are used to highlight the layer structure: vertices within an oval do not have connections, while vertices within a box connect to each other; an thick edge goes from one group to the other connects every vertex in one group to all the vertices in the other group.
The parameter mismatch pattern here is prescribed to coincide with the type of vertices, which is, for the given $\epsilon$: 
$[-\epsilon,-\epsilon,-\epsilon,+\epsilon,+\epsilon,+\epsilon]$. 
}
\label{seqnet}
\end{figure}

\begin{figure}
\centering
\includegraphics*[width=0.8\textwidth]{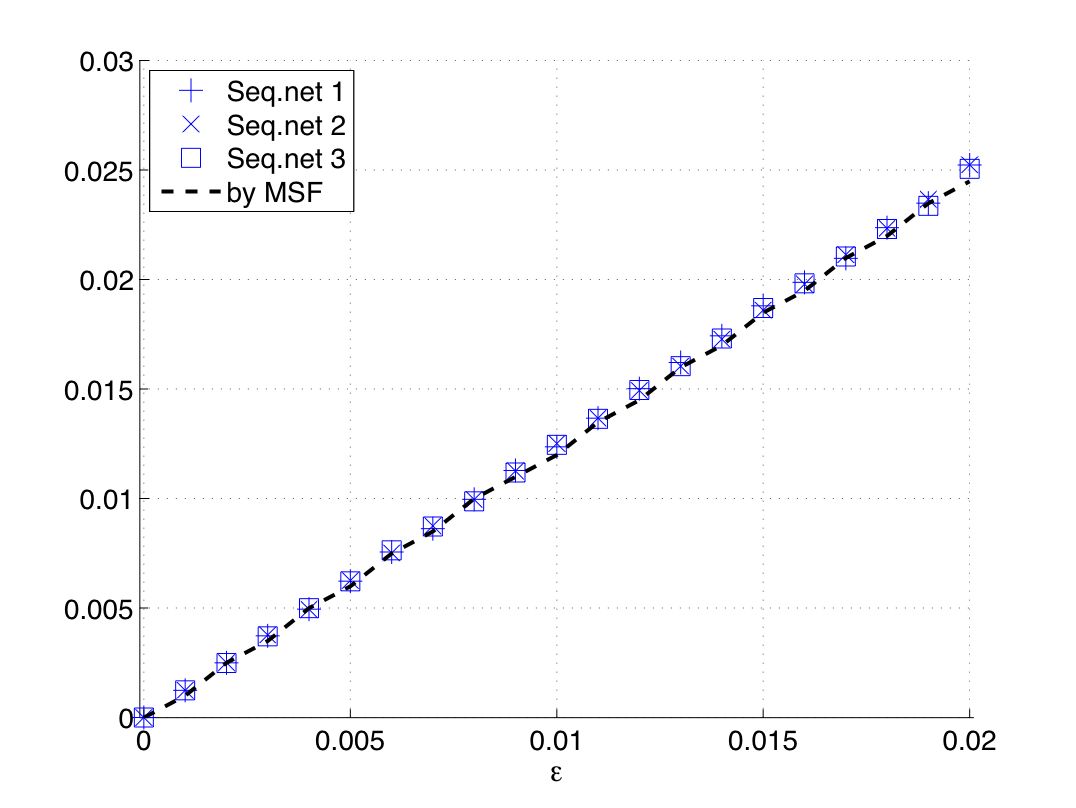}
\caption{(Color online) Validating MSF on sequence networks. Here the coupling strength is $g=2$. The parameter mismatch pattern is shown in Fig.~\ref{seqnet}. The vertical axis represents the average variation at each given $\epsilon$. (Blue) markers represent variations obtained from actual time series, and (black) dashed lines is the prediction obtained by MSF. Here the MSF line for all three networks are the same.}
\label{seqvar}
\end{figure}

Sequence networks\cite{SUN_PRE08} are a special class of networks that can be encoded by the so called {\it creation sequence}. In Fig.~\ref{seqnet} three different sequence networks of the creation sequence $(A,A,A,B,B,B)$ under different connection rules are shown. Interestingly, despite the fact that the structure of these networks are different, the variation of individual units to the average trajectory are the same, under the mismatch pattern 
$[-\epsilon,-\epsilon,-\epsilon,+\epsilon,+\epsilon,+\epsilon]$, see Fig.~\ref{seqvar}. 

Study on the eigenvector structure on these networks shows that this comes from the fact that the eigenvectors of all these three networks are the same, and more importantly, the parameter mismatch vector 
$[-\epsilon,-\epsilon,-\epsilon,+\epsilon,+\epsilon,+\epsilon]$
is parallel to one of the eigenvectors, corresponding to the same eigenvalue $\lambda=6$ in all three cases. Thus, the only active error mode in the eigenvector basis are the same for all three networks, resulting in the same variations.

\subsection{Example: Scale-free Networks}
The synchronization stability of a large network, with the knowledge of the probability distribution of parameters, is another interesting problem. To show how an expected MSF will apply, we use a scale-free network as an example. The network is generated using the {\it BA model}\cite{BARABASI_SCIENCE99}: start with a small initial network, consecutively add new vertices into the current network; when a new vertex is introduced, it connects to $m$ preexisting vertices, based on the {\it preferential attachment} rule\cite{BARABASI_SCIENCE99}. The network generated through process is known as a {\it BA network}, which is one example of a scale-free network. Here we use generate such a BA network with $N=500$ vertices and $m=12$.

\begin{figure}
\centering
\includegraphics*[width=0.8\textwidth]{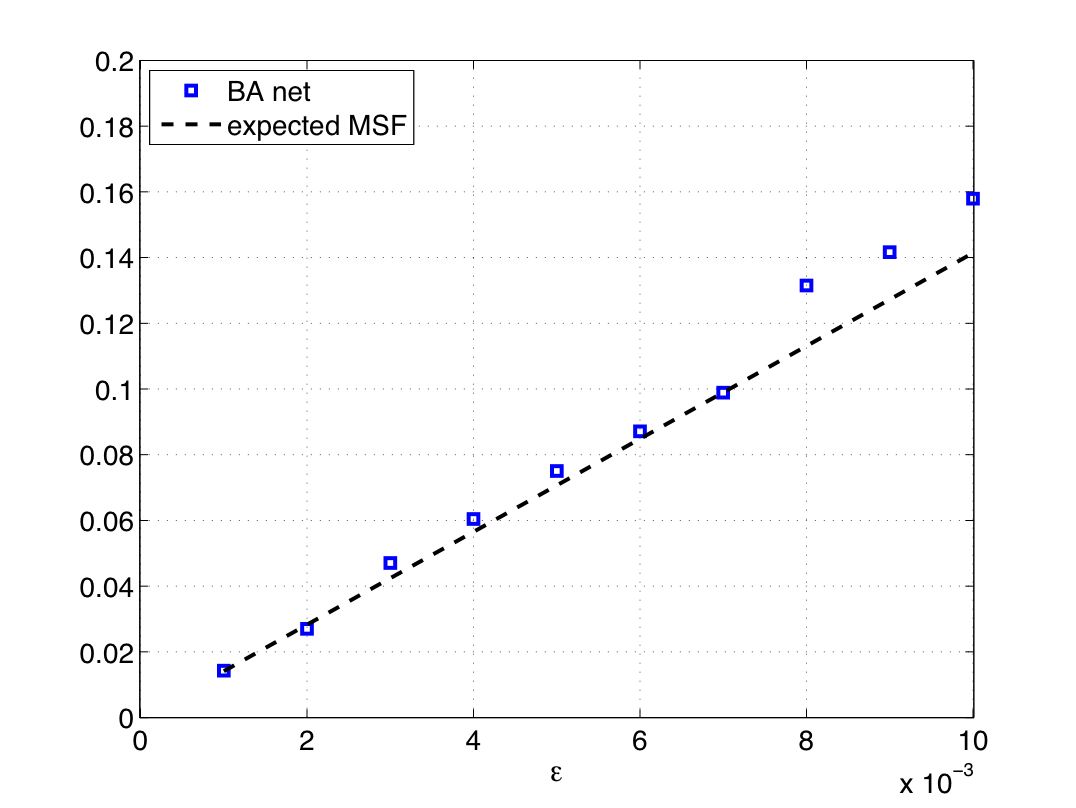}
\caption{(Color online) Validating MSF prediction for a BA scale-free network. The network consists of $500$ vertices with average degree $12$. Parameter on unit $i$ follows a Gaussian distribution $N(28,\epsilon^2)$. The expected MSF is obtained through $\Omega(g\lambda_i,\epsilon)$.}
\label{BAvar}
\end{figure}

In Fig.~\ref{BAvar} we show how parameter mismatch affect synchronization on a BA network. The parameters on each unit are assumed to follow the Gaussian distribution with mean $28$ and standard deviation $\epsilon$ for each given $\epsilon$. The expected MSF, as described in Sec. 3.1, predicts well the actual variation to the average trajectory, see Fig.~\ref{BAvar}.
	
\section{Summary}
In this paper we analyze the synchronization stability for coupled near-identical oscillator networks such as Eq.~\ref{origdyn}. We show that the master stability equations and functions can be extended to this general case as to analyze the synchronization stability. The variational equations in the near-identical oscillator case highlight the relevance of eigenvectors as well as eigenvalues on the effect of parameter mismatch, which indicates the importance of knowledge of the detailed network structure in designing dynamical systems that are more reliable.

\section{Acknowledgments}
J.S. and E.M.B have been supported for this work by the Army Research Office grant 51950-MA.

%

\end{document}